# Nanostructural insights into the dissolution behavior of Sr-doped hydroxyapatite

Hui Zhu[1,2], Dagang Guo[1,*] , Lijuan Sun[1], Hongyuan Li[1], Dorian A. H. Hanaor[2], Franziska Schmidt[2], Kewei Xu [1,3].

[1] State Key Laboratory for Mechanical Behavior of Materials, School of Material Science and Engineering, Xi'an Jiaotong University, Xi'an 710049, PR China

[2] Fachgebiet Keramische Werkstoffe/Chair of Advanced Ceramic Materials, Institut für Werkstoffwissenschaften und -technologien, Technische Universität Berlin, Hardenbergstraße 40, 10623 Berlin, Germany

3    Xi'an University of Arts and Science, Xi'an 710065, PR China

*Corresponding Author: Dr. DaGang Guo.

Post: State Key Laboratory for Mechanical Behavior of Materials, School of Material Science and
      Engineering, Xi'an Jiaotong University, Xi'an 710049, PR China.

Email: guodagang@mail.xjtu.edu.cn

Tel.: +86 29 82668614

Fax: +86 29 82663453

ABSTRACT

In this study, high-resolution transmission electron microscopy (HRTEM) was employed to characterize the nanostructure of strontium-substituted hydroxyapatite (Sr-HA) and its evolution following in vitro immersion in physiological solutions. HRTEM images showed that the substitution of Sr induced local distortions in the hydroxyapatite (HA) lattice: minor levels of edge dislocations were detected at low doping contents of Sr ions (1 at %); when the Sr content exceeded 10 at%, the density of grain boundaries increased notably and triple junctions were clearly observed. The dissolution of undoped HA was initiated at crystallite surfaces, whereas the dissolution of Sr-HA started around grain boundaries. Acicular nanocrystal reprecipitation was observed on grain surfaces immersed in simulated body fluid (SBF), while not in dilute hydrochloric acid (HCl). These findings suggest appropriate levels of Sr incorporation can introduce imperfections in the crystal structure of apatite and thus enhance its dissolution rate towards enhanced physicochemical performance in biomedical applications.

https://doi.org/10.1016/j.jeurceramsoc.2018.07.056

## 1. Introduction

Of all the calcium phosphate ceramics used as bone replacement materials, hydroxyapatite (HA; $Ca_{10}(PO_4)_6(OH)_2$) most closely resembles the main inorganic phase of bone and teeth in humans. It possesses a hexagonal lattice structure and a P63/m space group with cell dimensions of a=9.4211 and c=6.8810 Å [1], and is known to be outstandingly bioactive [2]. HA has been used in a range of medical applications such as coatings for implants and bone grafting materials. Bioactivity and biodegradability are generally accepted to be the two most critical features for third-generation biomedical materials [3]. However, it has been reported that the dissolution rate of bioactive HA in vivo is below 10 wt % per year, which is far slower than the growth rate of newly formed bone [4, 5]. Hence, numerous studies in recent years have aimed to improve the degradation rate of these materials. It has been found that the degradability of HA can be improved by various approaches such as tailoring its composition, crystallinity, surface area and Ca/P ratios [6-9]. Among such methods, ion substitution is particularly effectual because not only can it alter dissolution kinetics, but also may impart additional biological functionality mediated by the dopant ions themselves. For example, in separate studies, the substitution of carbonate, strontium, magnesium and silicon ions at Ca or P sites in HA has been shown to enhance degradation kinetics while the release of these cations further promotes cell growth, thus enhancing the interface between the bioceramic and biological tissue [10-12]. Porter et al. successfully utilised High Resolution Transmission Electron Microscopy (HRTEM) to demonstrate the superior solubility of silicon-substituted HA (Si-HA) compared to phase-pure HA [13, 14]. These improved dissolution kinetics can be attributed to the higher density of dislocations, grain boundaries and triple junctions in the doped HA structure, which are found to act as dissolution initiation sites in vivo [13, 15].

Owing to its chemical similarity to Ca, strontium (Sr) is a logical candidate for substitution in HA. Sr substituted HA materials have been previously studied with results showing enhanced properties in terms of biocompatibility [16, 17], osteoconduction [16], bioactivity [18, 19], mechanical performance [17] and degradation rate [20]. Notably, it was found that the mean weight loss of 10 at% strontium-substituted HA (Sr-HA) cement increased by 73.9 % in comparison with that of undoped material [21]. In addition, low doses of Sr are found to play important roles in the formation of new bone [22, 23], the restraint of bone resorption [23, 24], the treatment of bone cancer and the relief of bone pain [25], further motivating its incorporation in HA systems. Sila-Asna et al. proposed that the optimal concentration of strontium release was between 0.2107 and 21.07 μg/ml, whereas higher Sr concentrations up to 210.7 μg/ml may retard osteoblastic differentiation through a negative feedback mechanism [26, 27]. In addition, an animal experiment also showed that when fed in large amounts, strontium may cause rickets in experimental animals by disrupting the intestinal calcium absorption, synthesis of vitamin D and mineralization [9].

Overall, the growing volume of data regarding the beneficial effects of Sr ions on bone stimulates further research of Sr-HA systems. So far, however, research into Sr-HA has focused on materials synthesis, composition design and performance evaluation. Little attention has been paid to interpreting the mechanisms through which *in vivo* biodegradation of Sr-HA is enhanced relative to pure HA. Although it has been speculated that the enhancement of the dissolution rate arises due to variations in the crystallography of Sr-HA [28], there is a lack of experimental evidence in this regard. In particular the mechanisms through which Sr substitution alters the defect structure and consequent degradation of HA have not been clarified. These aspects are significant in order to optimize the composition of Sr-HA biomaterials.

Numerous studies have shown that the chemical dissolution of bone implant materials is the decisive factor in their in *vivo* degradation [29, 30]. Therefore, in this work, *in vitro* dissolution assays were carried out to simulate the performance of applied HA bioceramics. The present work has three principal aims; (i) to characterize the nanostructural features of Sr-HA with different Sr contents before and after *in vitro* immersion in physiological solutions by using HRTEM; (ii) to examine the relationships between dissolution characteristics, defect structure and Sr content in Sr-HA systems; (iii) to clarify the mechanism through which Sr substitution affects the degradation of HA.

## 2. Experimental section: materials and methods

### 2.1. Processing of strontium-substituted hydroxyapatite

Sr-HA materials having the stoichiometry $Ca_{10-x}Sr_x(PO_4)_6(OH)_2$ with x=0.1, 1 and 2, are referred to as 1 at% Sr-HA, 10 at% Sr-HA and 20 at% Sr-HA respectively. These materials were synthesized by a wet chemical method with the following steps [31]: (1) Ca-/Sr-containing solution with certain concentration was prepared by dissolving calcium nitrate $(Ca(NO_3)_2 \cdot 4H_2O$, analytically pure grade) and strontium nitrate $(Sr(NO_3)_2,$ analytically pure grade) in deionized water. The pH value of the mixed solution was adjusted to 10.0 by an aqueous ammonia solution. (2) A di-ammonium hydrogen phosphate $((NH_4)_2HPO_4$, analytically pure grade) solution was prepared by dissolving this salt in deionized water and then adjusting the pH to 9.0 with ammonia. (3) The di-ammonium hydrogen phosphate solution was slowly added to the mixed solution of calcium nitrate and strontium nitrate following the molar ratio of (Ca+Sr)/P =1.67 accompanied by constant stirring. The solution was carefully sealed after the pH value was adjusted to 10.5. (4) The reaction mixture was loaded in a conical flask and vibrated in a water-bath at 50°C at a rate of 130 rpm for 48h. (5) The reaction product was statically aged in the water-bath for 24 h. (6) After aging, the product was repeatedly separated by centrifugation and washed, four times with deionized water and finally with absolute ethyl alcohol. (7) The obtained filter cake was then dried in a vacuum oven at 100 ℃ (air-pressure< 0.1 Pa) for at least 10 h. (8) The dried bulk was crushed and ground manually and was then put into the vacuum oven (air-pressure<0.01 Pa) at 900 ℃ for 2 h. (9) Finally, the heat-treated powder was ground again and passed through a 200 mesh sieve.

In this study, Sr content in the final Sr-HA product was regulated by changing the ratio of calcium nitrate to strontium nitrate in the initial reaction mixture described in the above step (1). For HRTEM analysis, powders were ultrasonically dispersed in ethanol at a concentration of 1 mg/ml and deposited on 300-mesh copper grids. Before measurements, the samples were vacuum-dried at 37 ℃ for 24 h and sealed.

## 2.2. Dissolution studies

To replicate the performance environment of an applied bone implant material we prepared simulated body fluid (SBF), containing key ions in concentrations following the standard method of Kokubo [REF]. To achieve this formulation, SBF was prepared by dissolving reagent-grade NaCl, KCl, $K_2HPO_4 \cdot 3H_2O$, $MgCl_2 \cdot H_2O$, $CaCl_2$, and $NaSO_4$ in distilled water and buffering to pH 7.4 with tris [hydroxymethyl] aminomethane [$(CH_2OH_3)_3CNH_3$] and hydrochloric acid. Pure HA and 10 at% Sr-HA materials in powder form were individually immersed in SBF at 37 ℃ for 3 hours (h) and 3 days (d), respectively. In addition to SBF immersions, diluted hydrochloric acid was used in studies to accelerate the dissolution of the calcium phosphate salts.

## 2.3. HRTEM studies of samples before and after in vitro immersion in physiological solution

HRTEM, selected area electron diffraction, convergent beam electron diffraction, and energy dispersive X-ray spectroscopy (EDX) were performed on a JEM-2100F High Resolution Transmission Electron Microscope, which was operated at 200 kV with a point-to-point TEM resolution of 0.23 nm. Fast Fourier transformation (FFT) analysis and inverse FFT analysis were performed using DigitalMicrograph™ software (Gatan Inc.) to characterize the crystal structural variations in the samples.

## 3. Results

### 3.1. Variations in crystal structures of HA and Sr-HA

Sr-HA compositions of the general formula $Ca_{(10-x)}Sr_x(PO_4)_6(OH)_2$, where x=0.1, 1, 2, were studied alongside pure HA material (x=0). Well defined nano-crystallites of pure HA are shown in the TEM image of Fig.1a. It was observed that atoms inside the single crystal of pure HA had a periodic and highly ordered arrangement. For 1 at% Sr-HA, the orientation and the arrangement of the atoms remained well ordered, only with a few minor dislocations revealed by FFT analysis (Fig. 1b, c). At an increased Sr content of 10 at%, obvious lattice distortion and dislocations were apparent on the dominant (100) faces, which were indexed on the basis of their lattice spacing of 0.8168 nm (Fig.1d, e). Dislocations were mainly distributed near grain surfaces rather than internally. In addition, a higher density of grain boundaries was present, as shown in Fig. 1f. Notably in Fig. 1g, more triple junctions could be detected in 10 at% Sr-HA as well. For 20 at% Sr content, the major structure defects in Sr-HA were grain and subgrain boundaries (Fig. 1h). The areal density of observed grain boundaries increased with the growing Sr content. In Fig. 1i, a typical triple junction in 20 at% Sr-HA with obvious lattice distortion is shown. Overall, increasing dopant levels in apatite produced a significant effect on the apatite structure and induced crystal distortions and defects.

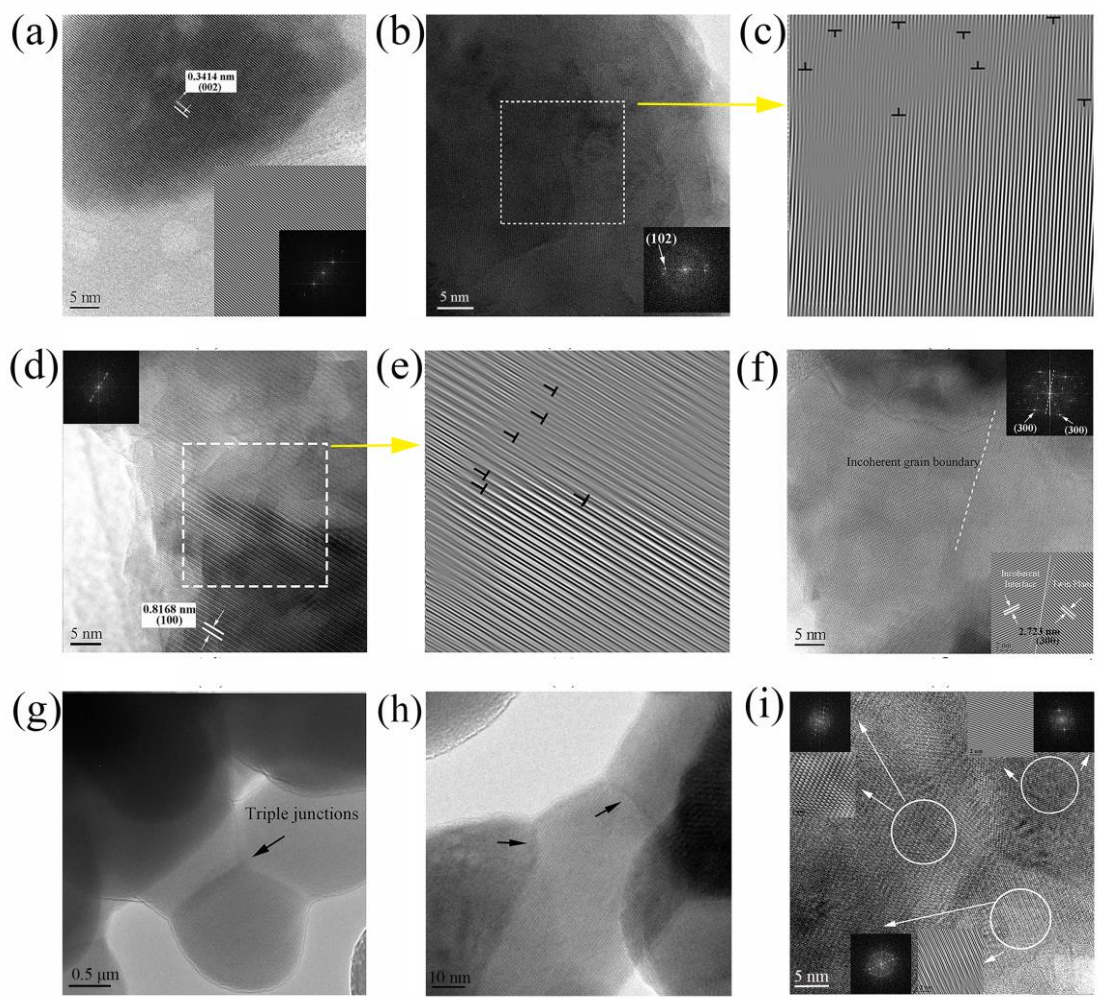

**Fig. 1.** High-resolution lattice images of pure HA and Sr-HA: (a) highly-ordered arrangement of pure HA; (b) a few small dislocations in 1 at% Sr-HA and (c) its FFT analysis; (d) ) dislocations and distortions on large planes (100) in 10 at% Sr-HA crystals, and (e) its FFT analysis; (f) incoherent twin boundary with similar interplanar distance in 10 at% Sr-HA; (g) triple junctions in 10 at% Sr-HA; (h) grain boundaries in 20 at% Sr-HA; (i) triple junctions in 20 at% Sr-HA

## 3.2. Dissolution characteristics of pure HA and 10 at% Sr-HA immersed in SBF

Here we examine 10 at% Sr-HA as the research object considering its evident nanostructural variations and biological utility. Different nanostructures and morphologies of pure HA and 10 at% Sr-HA were observed following immersion in SBF, which was conducted for durations of either 3 hours or 3 days.

After a 3h-immersion in SBF, pits appeared on the surface of pure HA grains, giving a mottled appearance evident of dissolution (Fig. 2a). After being immersed for 3 days, HA exhibited crystallites of two different forms, as shown in Fig. 2b: the original apatite was present as larger crystallites of irregular spherical form in the range of 100~200 nm, while newly formed finer acicular nanocrystals appeared near the surface of the original grains, with dimensions of approximately 5 nm in diameter and 12 nm in length, which were shown with a higher magnification in the insets of Fig. 2b. Furthermore, the presence of voids on the grain surface indicated that extensive dissolution of HA had occurred (Fig. 2b). Following the longer dissolution period of 3 days, a layered structure with irregular surfaces was observed for pure HA grains (Fig. c, d), suggesting a dissolution process with a lamellar-like peeling mode.

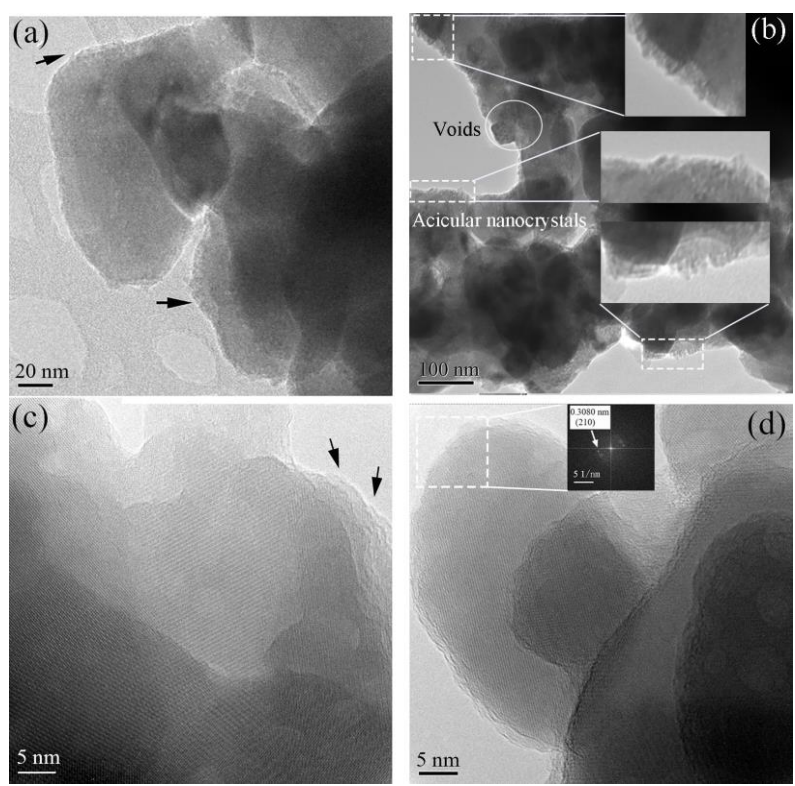

**Fig. 2.** (a) TEM micrographs of the pure HA grains after 3 hour immersion in SBF. The mottled appearance of grains suggested the occurrence of dissolution (arrows). Pure HA immersion in SBF after 3 days, illustrating (b) the emergence of small acicular nanocrystals (inset pictures) from the surface of the original apatite grains, which is related to the extensive dissolution, and (c-d) the extensive dissolution of pure HA extending along the grain surface with a lamellar-like peeling mode.

In contrast to pure HA, after 3 hours of immersion in SBF 10 at% Sr doped materials showed dissolution starting from exposed grain boundaries as shown in Fig.3a. After 3 days, the degree of dissolution increased, accompanied by the reprecipitation and growth of numerous elongated or needle-like crystallites, exhibiting dimensions of about 5 nm in diameter and 14 nm in length, as well as voids on the surface of larger HA grains (diameters 60~150 nm) (Fig. 3b). It is
.

worth noting that these crystallites formed in an acicular growth habit and were almost perpendicular to the original HA surface. A higher magnification image given in Fig. 3c shows that the grain boundary of the (210) plane was the initiation point for dissolution in the doped material, and further shows the formation of thin amorphous surface layers as confirmed by selected area electron diffraction (SAED)

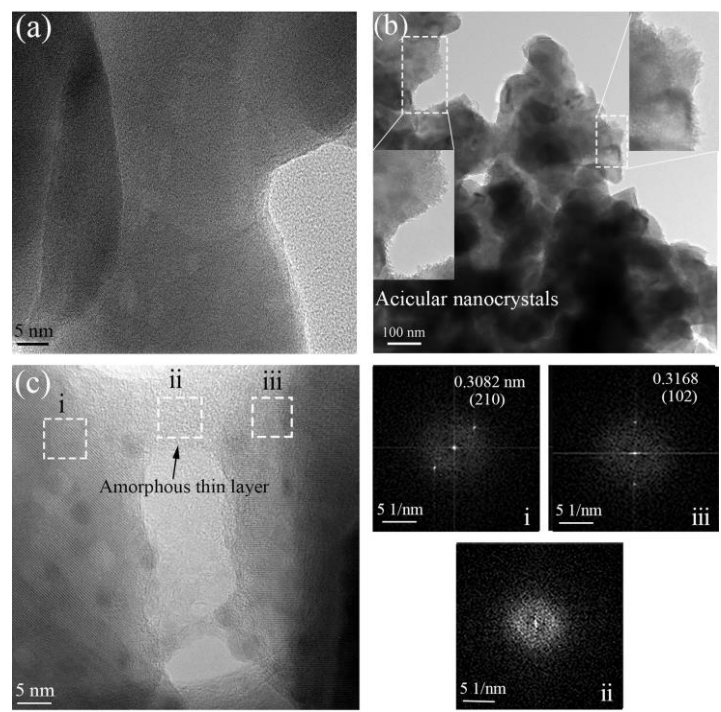

**Fig. 3.** TEM micrographs of 10 at% Sr-HA immersed in SBF (a) after 3 h, indicating the preference of Sr-HA dissolution at the grain edges and the grain junctions. (b) after 3 days immersion in SBF, illustrating extensive dissolution accompanied by acicular crystallites (inset pictures) and voids emanating from the grain surface; (c) dissolved regions around the grain boundary following a 3 day SBF immersion, with SAED patterns shown to the right confirming the formation of a thin amorphous layer.

Overall, the dissolution of 10 at% Sr-HA started mainly from the intersection of grain boundaries with grain surfaces and then extended along the grain boundaries. As the dissolution proceeded amorphous thin layers formed at the dissolution front. Edge dislocations in external crystallites was also identified as an initiation site for surface dissolution.

### 3.3. Immersion in diluted hydrochloric acid (HCl)

To enhance the influence of crystal defects on HA dissolution, another group of HA and Sr-HA samples were immersed in diluted HCl solution instead of SBF (see Fig. 4 and Fig. 5, respectively). It can be seen that the dissolution rate of both materials was faster in HCl solution than in SBF.

In general, the dissolution of pure HA was more pronounced in HCl solution than in SBF at surface regions following a 3-hour immersion (Fig. 4a). After 3 days, crystals with irregular spherical structures (50 nm~200 nm) and typical etch pits with crenulated fringes formed from the rapid dissolution in HCl. Notably, HA grain surfaces showed extensively mottled morphology with numerous voids (Fig. 4b) after 3 days' immersion. The dissolution process continued in the vicinity of those mottled structures on grain surfaces (indicated by a black arrow) (Fig. 4c). While the internal crystal structure was still intact, showing no signs of dissolution (Fig. 4d).

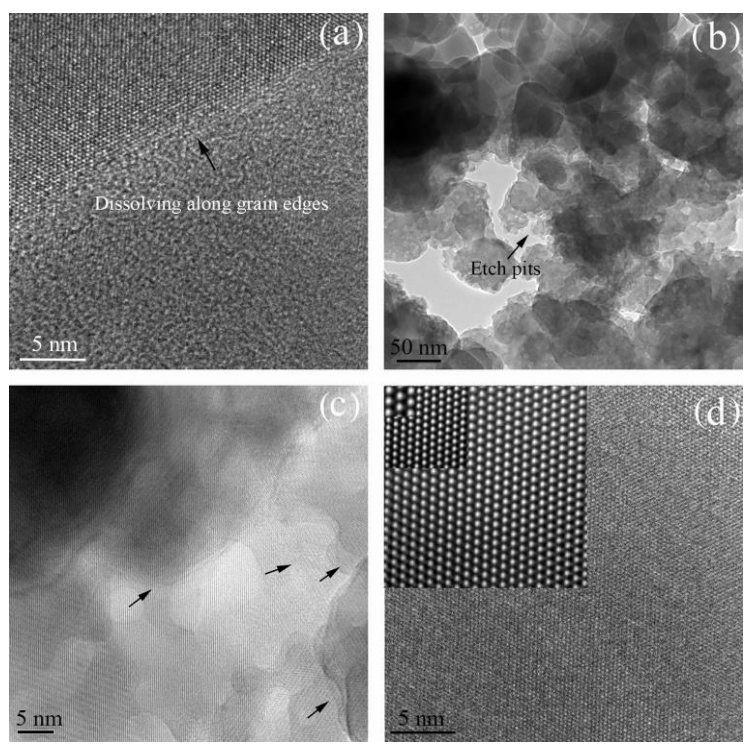

**Fig. 4.** (a) TEM micrographs of pure HA grains after 3 hour immersion in HCl solution. Black arrows show dissolution that occurred predominantly at grain surfaces; (b) mottled structures comprising numerous voids in pure HA after immersion in HCl solution for 3 days; (c) continuous dissolution propagating in the vicinity of mottled structures on grain surfaces; (d) High magnification TEM image of pure HA after 3 day immersion in HCl, showing no signs of dissolution inside the crystallite.

As can be seen in Fig. 5, the dissolution of HA in HCl solution increased dramatically with Sr content. We investigated a shorter immersion duration of Sr-HA in HCl based on its pronounced dissolution rate in Fig. 5. Within 5 minutes, dissolution had occurred in 10 at% Sr-HA, as shown in

Fig. 5a. Dissolution initiated preferentially from intersections between grains, including grain boundaries and triple junctions. Additionally, the initiation of dissolution at grain surfaces was also observed (Fig. 5b). After a 3 h immersion, numerous large voids and pits were found on the mottled structure of the original HA surfaces (Fig. 5c), which indicates a deeper level of dissolution. Besides, the average size of the 10 at% Sr-HA grains with spherical structures of 50-100 nm was smaller than that of pure HA after 3 h immersion. As shown in Fig. 5d-g, dissolution expanded around triple junctions, grain boundaries and surfaces of 10 at% Sr-HA after 3 h immersion. SAED pattern revealed that the crystalline structure at the dissolution forefront around the grain boundaries and triple junctions transformed into amorphous thin layers (Fig. 5d, e).

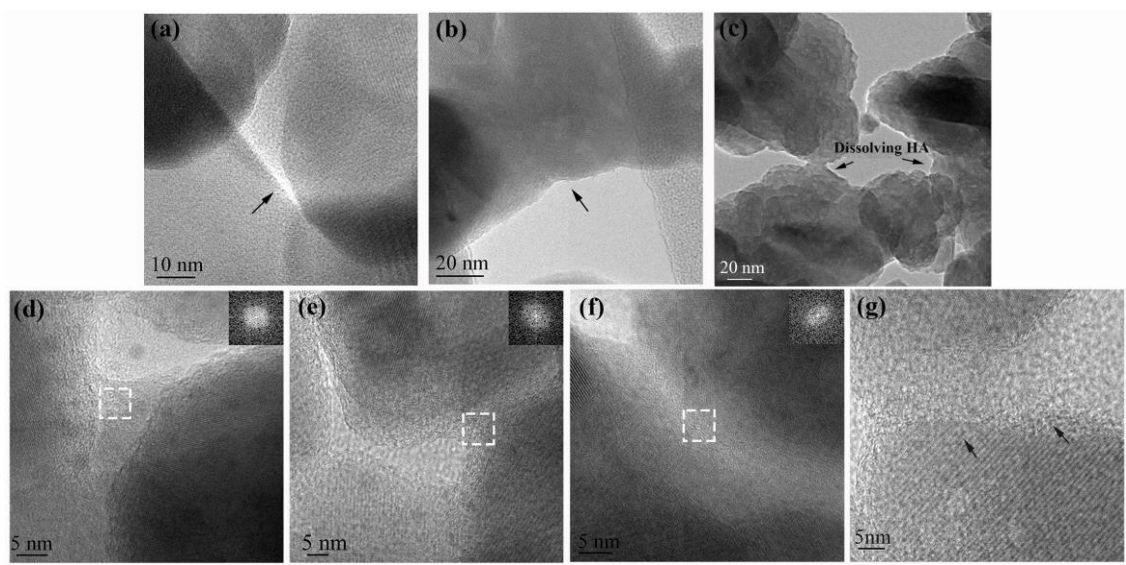

**Fig. 5.** TEM micrographs of the dissolution of 10 at% Sr-HA in HCl solution for 5 min, indicating (a) preferential dissolution at intersections, including grain boundaries and triple junctions; (b) dissolution at grain surfaces. 10 at% Sr-HA after 3 hours immersion in HCl solution showed (c) a greater depth of dissolution accompanied by the formation of a number of large voids on the grain surface; (d) and (e) extensive dissolution around triple junctions and the transformation from crystal to amorphous structure at dissolution forefront; (f) extensive dissolution around grain boundaries and the transformation from crystalline to amorphous structure at the dissolution forefront; (g) continuous dissolution along grain surfaces and the transformation from crystal to amorphous structure at the dissolution forefront.

### 3.4. Chemical analysis before and after dissolution

In order to clarify the distribution of Sr ions inside the Sr-HA lattice, EDX was adopted in this work. Line and map scanning were applied respectively on the internal grain and grain boundary of 10 at% Sr-HA (see Fig. 6). The Ca/Sr ratios (at. %) of both regions are listed in Table 1. It is clear that the Ca/Sr ratio was 13.50 in the intracrystalline while 9.46 around the grain boundary. The difference in the Ca/Sr ratio indicates that a larger amount of Sr ions was located at grain boundaries.

**Fig. 6.** TEM images of typical 10 at% Sr-HA with a grain boundary. Rectangle and line represent the map scanning and line scanning for EDX, respectively.

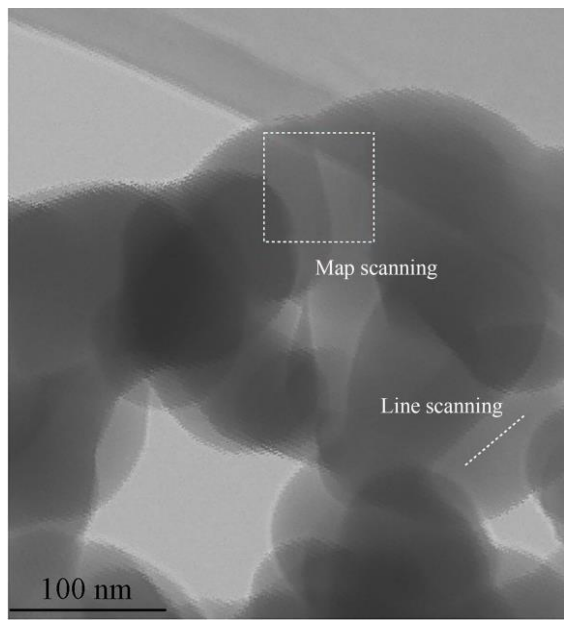

**Table 1.** The Ca/P ratio inside HA grain and on its grain boundary

| (a) | Element | Wt. % | At. % | (b) | Element | Wt. % | At. % |
|---|---|---|---|---|---|---|---|
| | O K | 41.26 | 62.50 | | O K | 34.86 | 56.56 |
| | P K | 20.86 | 16.33 | | P K | 21.62 | 18.12 |
| | Ca K | 32.60 | 19.72 | Grain | Ca K | 35.35 | 22.90 |
| Internal Grain | Sr L | 5.28 | 1.46 | Boundary | Sr L | 8.16 | 2.42 |
| | Resulting ratio of Ca/Sr | Theoretical Ratio of Ca/Sr | | | Resulting ratio of Ca/Sr | Theoretical Ratio of Ca/Sr | |
| | 13.50: 1 | 9: 1 | | | 9.46: 1 | 9: 1 | |

Subsequently, the ratio of Ca to Sr ions in apatite before and after immersion was analyzed. As is shown in Table 2, the resulting Ca/Sr ratio rose from 8.92:1 to 9.71:1 following immersion, indicating that the release rate of $Sr^{2+}$ was higher than of $Ca^{2+}$ ions in vitro.

**Table 2.** Relative Changes of the Composition of the 10 at% Sr-HA before and after dissolution

| [Sr]= 10% | Element | Wt. % | At. % | [Sr]= 10% | Element | Wt. (%) | At. % |
|---|---|---|---|---|---|---|---|
| Before dissolution | O K | 29.31 | 51.25 | After dissolution | O K | 45.45 | 58.28 |
| | P K | 16.75 | 15.12 | | P K | 13.56 | 8.98 |
| | Ca K | 43.34 | 30.25 | | Ca K | 23.73 | 12.14 |
| | Sr L | 10.61 | 3.39 | | Sr L | 5.33 | 1.25 |
| | Resulting ratio of Ca/Sr | Theoretical ratio of Ca/Sr | | | Resulting ratio of Ca/Sr | Theoretical ratio of Ca/Sr | |
| | 8.92: 1 | 9: 1 | | | 9.71: 1 | 9: 1 | |

## 4. Discussion

The current work examines the mechanisms through which Sr-induced defect structures affect the dissolution of HA. To this end, we first characterized defect structures including dislocations, grain boundaries and triple-junctions for different dopant concentrations in Sr-HA. Subsequently we investigated the dissolution morphology, initiation sites, and dissolution propagation for pure HA and 10 at% Sr-HA immersed in SBF and diluted hydrochloric acid.

Prior work has indicated that the substitutional incorporation of larger $Sr^{2+}$ ions into apatite leads to distortion of its hexagonal structure and an enhancement of its dissolution [10]. Li et al. demonstrated that changes in the crystallinity of synthetic apatite were governed by the concentration of the substitutional strontium dopant [32]. An X-ray diffraction analysis reported in our previous work [31] confirmed the shift of diffraction peaks to lower $2\theta$ angles in Sr-HA, indicating lattice expansion and a corresponding increase in the d-spacing between crystallographic planes, as summarised in Table 3. This is to be expected as $Sr^{2+}$ is slightly larger than $Ca^{2+}$ (118 and 100 pm, respectively, for octahedral coordination) [33], leading to an expansion in volume of the HA unit cell with increasing substitution levels.

**Table 3.** Structural and Profile Data for Pure HA,1,5,10,20,50,100 % Sr-HA [31]

| %Sr | 0 | 1 | 5 | 10 | 20 | 50 | 100 |
|---|---|---|---|---|---|---|---|
| a-Axis(Å) | 9.42 | 9.42 | 9.44 | 9.45 | 9.49 | 9.60 | 9.77 |
| c-Axis(Å) | 6.88 | 6.88 | 6.90 | 6.92 | 6.96 | 7.09 | 7.28 |

In the present work, at low levels of Sr dopant (1 at%), minor levels of dislocations were detected; for doping of 10 at%, numerous grain boundaries and triple junctions formed. This might be explained by the different sites where Sr ion is located in the HA lattice with varied Sr content. Two cation sites, namely Ca(I) and Ca(II), exist in the lattice structure of HA [31]. Rietveld analysis has shown that there is a clear preference of doping at site I for lower levels of Sr doping (1~5 at%); while for higher amounts of Sr in the crystal lattice, there is a slight preference of site II substitution [34]. The gradual doping into Ca-I sites results in minor or undistorted expansion of the crystal lattice, which can be associated with the uniform shift of the $PO_4$ tetrahedron surrounding the Ca-I site, towards neighboring hydroxyl channels, allowing for better accommodation of bigger ions of Sr at the Ca-I site. The few dislocations observed in 1 at% Sr-HA might be caused by the modest lattice expansion and associated voids. In contrast, the mixed doping or progressive substitution of Sr into Ca-II sites causes more extensive distortions of the unit cell [35]. This is in good agreement with the results of Terra et al.[34]. Based on those theoretical studies, when the Sr levels exceeds 10 at%, crystal twist and deformation appear to absorb the energy of vigorous distortion, resulting in shear defects such as grain boundaries and triple junctions.

Enhancing the *in vivo* degradation of bone implant materials is targeted as a key pathway towards improving their osseointegrative performance. This process involves chemical dissolution, body fluid erosion, cell phagocytosis and digestion. It is well accepted that the chemical dissolution properties of bone implant material are ultimately what determine its *in vivo* degradation rate [29, 30]. Based on this, *in vitro* dissolution was studied here to gain insights into the merit of Sr-HA bioceramics. We focused here on 10 at% Sr-HA due to its reportedly superior biological performance compared with other Sr dopant levels [36]. In a study by Sila-Asna et al. it was shown that for osteoblastic differentiation an optimal concentration of strontium ranelate lay in the range 0.2107 and 21.07 μg/ml, whereas higher concentrations bring about detrimental effect on this differentiation [26][18]. In addition, when consumed in large amounts, strontium was found to cause rickets in animals by disrupting intestinal calcium absorption and synthesis of vitamin D [9]. These studies imply that higher levels of Sr in HA would be undesirable, discouraging the investigation of materials with dopant levels over 10 at%.

The structure and composition of a material largely determine its dissolution performance. It has been proven that bone apatite dissolution processes initiate at crystal defect sites and further expand along dislocation lines [22]. For example, enamel apatite dissolves *in vivo* owing to the existence of point defect groups, dislocations, stacking faults and amorphous grain boundary phases [23, 24], with these defects related to the presence of trace elements [24]. *This is supported by further in vivo studies* [37, 38]. In agreement with these findings, here it was found that 10 at% Sr-HA exhibits loss of material mainly from defect structures, especially grain boundaries and triple junctions,. These findings were consistent for the enhanced dissolution studies performed in HCl solutions [39]. Defect structures are associated with structural disorder at the nanoscale. This leads to higher interfacial energy and relatively weaker bonding energies [40]. As a result, grain-boundary atoms more readily break away compared to the intracrystalline atoms, which favors the initiation of dissolution at grain boundaries and triple junctions in Sr substituted HA. This is supported by simulations that have shown grain boundary structures govern dissolution behavior of biological apatites [33]. Furthermore, in doped HA, with prolonged immersion, a thin amorphous layer is formed at the dissolution front. The formation of this amorphous layer might be interpreted as the result of an interfacial dissolution – reprecipitation process, which has been reported in previous research [41-43].

In contrast, grain boundary dissolution and amorphous layers were not observed in pristine HA in SBF and HCl. Instead, dissolution mainly emanated from the grain surface. Because the undoped HA lattice has few crystal defects, when the relaxation and reconstruction of the hydroxyapatite surface is sufficiently large to destroy the pairing of hydroxy groups, the OH⁻ groups are pulled apart and lose their pair-wise interaction, which induces the onset of surface dissolution via the release of hydroxy groups from the material [44]. After longer immersion time, the surface calcium ions also show the onset of dissolution, through lengthening of the bonds to lattice oxygen ions and solvation by the surface water molecules [44]. This surface dissolution extended showing a peeling layer morphology, indicating that the atoms in the apatite were gradually dissolved and peeled off from the pure HA surface.

A comparison between the dissolution characteristics of pure HA and Sr-HA in both SBF and HCl solution is provided in Table 4. The prominent grain boundary dissolution in Sr-HA can also be confirmed by a preferential release of strontium ions from Sr-HA. In our previous study, Rietveld structural refinements clearly illustrated that strontium ions substituted for calcium ions in the HA lattice [31]. Based on EDX analysis of the Ca/Sr ratios of internal and grain boundary regions in 10 at% Sr-HA (Table 1), it is postulated that strontium segregates as a strontium-rich phase at some grain boundaries in Sr-HA, possibly owing to a lower surface-free energy [45]. This study further found that the atomic ratio of Ca to Sr ions decreased after immersion, implying the preferential dissolution of strontium. This suggested that the increased concentration of Sr at grain boundaries regions may enhance the localised initiation of dissolution at these regions However, further studies are needed to ascertain the segregation behavior of Sr ions and coordination at grain boundaries.

**Table 4.** Universal dissolution grade of phase-pure HA and 10%Sr-HA in both SBF and diluted HCl solution

| | Immersion in SBF | | Immersion in diluted hydrochloric acid | |
| --- | --- | --- | --- | --- |
| | Pure HA | 10 at% Sr-HA | Pure HA | 10 % Sr-HA |
| Dissolution morphology | A mottled apperance with obvious pits on the original spherical structures(100~200 nm); acicular nanocrystals emanating from grain surface (20 nm diameter,100 nm length) with voids | Irregular spherical structures (60~150 nm) with the apperance of voids and acicular structures (20 nm diameter,70 nm length) that are perpendicular to original grain surface | A mottled apperance with numerous voids on the original spherical structures (50~200 nm), and etch-pit structures with crenulated fringes | A mottled apperance with large voids and pits on the original spherical structures (50~100 nm) |
| Starting site of dissolution | Grain surface | Grain boundaries and grain surface | Grain surface and the etch pit area | Grain surface and intersection among grains, including cross-boundaries and triple junctions |
| Dissolution propagation | Dissolution along the grain surface with a lamellar like peeling mode | Dissolution forefront translated to amorphous thin layer at grain boundaries | Dissolution along the grain surface in the vicinity of those etch pit structures | Dissolution along the defect structure in grains; the dissolution forefront transformed into amorphous thin layer |

In addition to segregation effects, simply the finer grain structure of substituted HA materials may enhance dissolution, as a decrease in crystallite is generally accepted to enhance solubility [14, 46, 47]. Interestingly, previous studies demonstrated that the crystallite size of HA decreased with Sr substitution levels up to 25 at% [46], which is in good agreement with our observations. This could present another potential pathway through which strontium ions increase the *in*

*vivo* bioactivity of HA.

Further to improving solubility, strontium doping might promote the heterogeneous nucleation of apatite, which is accepted as an essential requirement for the functional osseointegration of biomaterials [48]. *In vitro* immersion experiments using SBF are widely used to reproduce the formation of a biologically equivalent apatite layer in vivo. In the current study, acicular crystallites perpendicular to the grain surface were observed at the dissolution front of HA immersed in SBF. Based on earlier reports, such features are likely to be a reprecipitated bone-like apatite [49-51]. These elongated crystallites have a consistent size of 4-8 nm diameter and 12-15 nm length. The morphology of these features is shown by TEM and SEM micrographs in supplementary Fig. S-1. The reprecipitation of acicular apatite at surfaces of 10 at% Sr-HA was similar to that on pure HA (Figs. 2 and 3), but slightly larger crystals were observed on the Sr doped specimen, suggesting Sr substitution could enhance the osseointegrative functionality of HA biomaterial systems.

## 5. Conclusions

This study confirms that the substitution of calcium by strontium leads to structural disorder resulting in a loss of crystallinity and formation of various defects including dislocations, stacking faults, grain boundaries and triple junctions whose type and density varies with dopant level. As a consequence, the dissolution rate is enhanced with increasing Sr content. The sites where dissolution nucleates and the routes in which it expands in Sr-HA are quite different in comparison to pure HA: dissolution is initiated from crystal surfaces in pure HA, whereas dissolution initiation sites in Sr-HA are mainly concentrated around crystal defects and Sr-rich grain boundaries. The results of this study shed further light on the fundamental mechanisms of ion-doped apatite dissolution, with implications towards hard tissue repairing materials.

## Acknowledgements

This work was supported by the National Natural Science Foundation of China (No. 51072159; 51273159), Science and technology program of Shaanxi Province (No: 2014K10-07) and Technology Foundation for Selected Overseas Chinese Scholar, Department of Human Resources and Social Security of Shaanxi Province (No: 2014-27).